\pdfoutput=1
\documentclass{iopart} 
\usepackage{amssymb}
\usepackage{iopams} 
\usepackage{upgreek}
\usepackage{bm}
\usepackage{graphicx}
\eqnobysec
\begin{document}
\title{Self-force as probe of internal structure}     
\author{Soichiro Isoyama$^{1,2}$ and Eric Poisson$^1$} 
\address{$^1$ Department of Physics, University of Guelph, Guelph, Ontario,
N1G 2W1, Canada}
\address{$^2$ Yukawa Institute for Theoretical Physics, Kyoto university,
Kyoto, 606-8502, Japan}  
\ead{isoyama@yukawa.kyoto-u.ac.jp}
\ead{epoisson@uoguelph.ca} 
\date{April 16, 2012} 
\begin{abstract} 
The self-force acting on a (scalar or electric) charge held in place
outside a massive body contains information about the body's
composition, and can therefore be used as a probe of internal
structure. We explore this theme by computing the (scalar or 
electromagnetic) self-force when the body is a spherical ball of
perfect fluid in hydrostatic equilibrium, under the assumption that
its rest-mass density and pressure are related by a polytropic
equation of state. The body is strongly self-gravitating, and all
computations are performed in exact general relativity. The dependence
on internal structure is best revealed by expanding the self-force in
powers of $r_0^{-1}$, with $r_0$ denoting the radial position of the
charge outside the body. To the leading order, the self-force scales
as $r_0^{-3}$ and depends only on the square of the charge and the
body's mass; the leading self-force is universal. The dependence on
internal structure is seen at the next order, $r_0^{-5}$, through a
structure factor that depends on the equation of state. We compute
this structure factor for relativistic polytropes, and show that for a
fixed mass, it increases linearly with the body's radius in the case
of the scalar self-force, and quadratically with the body's radius in
the case of the electromagnetic self-force. In both cases we find that
for a fixed mass and radius, the self-force is smaller if the body is
more centrally dense, and larger if the mass density is more uniformly 
distributed.  
\end{abstract} 

\section{Introduction and summary} 
\label{sec:intro} 

An electric charge held in place in the curved spacetime of a massive  
body produces an electric field that responds to the spacetime
curvature; a consequence of the interaction is a distortion of the
field lines from an otherwise isotropic distribution near the charge,
which leads to a net force acting on the particle. This is the
physical origin of the electromagnetic self-force
\cite{dewitt-brehme:60}, a subtle effect that is sensitive to the
geometry of spacetime not just in the vicinity of the charge, but
everywhere.   

Self-force effects in curved spacetime have been vigourously explored; 
for an extensive review, see Ref.~\cite{poisson-pound-vega:11}. Most of
the recent activity was focused on the gravitational self-force, in an
effort to model the inspiral and gravitational-wave emissions of a
binary system with a small mass ratio \cite{barack-sago:10,
  diener-etal:11, warburton-etal:12}. The prototypical problem,
however, goes back to 1980, when Smith and Will \cite{smith-will:80}
calculated the self-force acting on a particle with electric charge
$e$ held in place at a radius $r_0$ in the Schwarzschild spacetime of
a nonrotating black hole of mass $M$. The analogous problem of the
self-force acting on a scalar charge $q$ in the same spacetime was
investigated by Wiseman \cite{wiseman:00}. The results are
interesting: The radial compoment of the electromagnetic self-force
(in Schwarzschild coordinates) is equal to $e^2 M f_0^{1/2}/r_0^3$, in
which $f_0 := 1-2M/r_0$, while the self-force vanishes in the case of
a scalar charge. The electromagnetic self-force is repulsive, meaning
that the external force required to keep the particle in place is
smaller when the particle is charged, compared to what it would be in
the case of a neutral particle. These results are not intuitive: It is
not obvious why the electromagnetic self-force should be repulsive,
and it is not obvious why the scalar self-force should vanish; for an
intriguing explanation based on Newton's third law, refer to Sec.~IV
of Ref.~\cite{burko-etal:01}.  

The issue that interests us in this paper is the way in which the
internal composition of the massive body affects the self-force acting
on an electric or scalar charge. Because the self-force is the result
of an interaction between the field and the spacetime curvature, and
because this interaction will depend on the details of internal
structure, it is expected that the self-force will reflect these
details, and will therefore reveal some aspects of the body's
composition. Our goal is to describe which aspects of the internal 
structure can be revealed by the self-force. 

This line of enquiry was initiated by Burko, Liu, and Soen
\cite{burko-etal:01}. Building on earlier work by Unruh
\cite{unruh:76}, they calculated the self-force acting on electric and
scalar charges held in place in the exterior spacetime of a massive
thin shell. They found that their results agreed with the black-hole
results to leading order in an expansion of the self-force in powers
of $r_0^{-1}$, but that structure-dependent terms appeared at
higher-order. The cases of an electric charge outside a body of 
constant density, and of an electric charge outside a massive
conductor, were examined by Shankar and Whiting
\cite{shankar-whiting:07}, who confirmed the agreement with the
black-hole results for large values of $r_0$. 

The enquiry was recently pursued by Drivas and Gralla
\cite{drivas-gralla:11}, who considered static charges in the exterior
spacetime of a massive body of arbitrary composition. They firmly
established the universality of the self-force at order $r_0^{-3}$,
and revealed a dependence on internal structure at order $r_0^{-5}$
and beyond. More precisely stated, Drivas and Gralla found that when
the self-force is expanded in powers of $r_0^{-1}$, the leading term
that appears at order $r_0^{-3}$ is independent of the body's
internal structure. In the case of an electric charge, it is always
equal to the Smith-Will result $e^2 M/r_0^3$, and in the case of a
scalar charge, it is always equal to the vanishing Wiseman
result. (This last statement holds for a scalar field that is
minimally coupled to the spacetime curvature. For nonminimal coupling,
the self-force continues to vanish when the charge is held outside a
black hole, but it is equal to $2\xi q^2 M/r_0^{3}$ when there is a
material body, where $\xi$ is a dimensionless coupling constant. In
this case the universality of the scalar self-force at order
$r_0^{-3}$ partially breaks down: the self-force is only insensitive
to the details of internal structure when the body is not a black
hole.)    

The main concern of Drivas and Gralla was to display the universality 
of the self-force at order $r_0^{-3}$, and their other discovery, that
there is a dependence on internal structure at order $r_0^{-5}$ and
beyond, was left largely unexplored. This is the theme that we intend
to pursue here: How does the self-force depend on the body's internal
structure, and which aspects of the internal composition can be
inferred from a close examination of the self-force? These issues are 
interesting, because it is very difficult, in general relativity, to
collect any information about a body's internal structure when the
body is spherical, and when one is limited to external
measurements. It is impossible, for example, to reveal any aspect of
the internal structure by measuring the motion of a test body. Going 
beyond test bodies, for example by measuring the orbital motion of a
two-body system of comparable masses, reveals very little: the strong
formulation of the principle of equivalence implies that the orbital
motion is necessarily insensitive to the details of internal
structure, until the bodies are so close together that significant
tidal deformations are generated \cite{flanagan-hinderer:08,
  read-etal:09, hinderer-etal:10, pannarale-etal:11,
  damour-nagar-loic:12, lackey-etal:12}. 

The self-force, on the other hand, can be used as a probe of internal
structure. Indeed, one can easily imagine a thought experiment
designed to exploit the self-force in this manner. Suppose that one is
given a neutral particle, a second particle with an electric charge
$e$ (or a scalar charge $q$), and a massless rope to keep each
particle in place in the body's gravitational field; the other end of
the rope is held at a safe distance by an external agent. Keeping the
neutral particle at a position $r_0$ requires the external agent to
hold the rope firmly, and the tension in the rope reveals the body's
gravitational force on the particle. Replacing the neutral particle by
the charge produces a slight deficit in the tension, which corresponds
to the repulsive action of the self-force; the self-force at position
$r_0$ can thus be measured. Repeating the measurement for many
positions allows one to infer the self-force as a function of $r_0$,
together with its expansion in powers of $r_0^{-1}$; the terms of
order $r_0^{-5}$ and beyond are those that reveal aspects of the
body's internal stucture. There is, of course, nothing practical about
this. But it is of interest to observe that as a matter of principle,
one can infer elusive details of internal structure by careful
measurements of a self-force acting on a static charge.   

To investigate the dependence of the self-force on the internal
structure of a massive body, we consider a spherical star that
consists of a perfect fluid in hydrostatic equilibrium. The fluid is
characterized by a (baryonic) rest-mass density $\rho$ and a pressure
$p$, and these variables are taken to be related by the polytropic
equation of state $p = K \rho^{1+1/n}$, in which $K$ is a scaling
constant, and $n$ is the polytropic index (another constant). The star
possesses a mass $M$ and a radius $R$, and a (scalar or electric)
charge is placed at a radius $r_0 > R$ in the vacuum region exterior
to the star.

Following Drivas and Gralla \cite{drivas-gralla:11}, we compute the
difference between two self-forces. The first is the actual self-force
acting on the charge held in place outside the polytropic star, and
the second is the self-force that would be acting if the charge were
instead situated outside a Schwarzschild black hole; the bodies have
the same mass, and the charge is placed at the same position outside 
each body. It is this difference that contains the dependence of the
self-force on the body's internal structure. We show below that in the
case of a scalar charge, the difference can be expressed as  
\begin{equation} 
\fl
\Delta F^r_{\rm scalar} = -\Bigl( \frac{q}{M} \Bigr)^2 
\biggl( \frac{z_0-1}{z_0+1} \biggr)^{3/2} \sum_{\ell=1}^\infty 
(2\ell + 1) S^{\rm scalar}_\ell Q_\ell(z_0) Q'_\ell(z_0),   
\label{intro:scalar_force} 
\end{equation} 
in which $z_0 := r_0/M - 1$, $Q_\ell(z_0)$ is a Legendre function of
the second kind, a prime indicates differentiation with respect to
$z_0$, and $S^{\rm scalar}_\ell$ is a structure factor
that depends on the stellar interior; an expression is provided below
in Eq.~(\ref{scalar_structure}). In the case of an electric charge we
have instead  
\begin{equation} 
\fl
\Delta F^r_{\rm em} = -\Bigl( \frac{e}{M} \Bigr)^2 
\biggl( \frac{z_0-1}{z_0+1} \biggr)^{3/2} \sum_{\ell=1}^\infty 
(2\ell + 1) S^{\rm em}_\ell \biggl[ Q_\ell(z_0) 
- \frac{(z_0-1)Q'_\ell(z_0)}{\ell(\ell+1)} \biggr] 
Q'_\ell(z_0), 
\label{intro:em_force} 
\end{equation} 
in which $S^{\rm em}_\ell$ is the corresponding structure factor for
the electromagnetic self-force; see Eq.~(\ref{em_structure}) below. 

\begin{figure}
\includegraphics[width=5in]{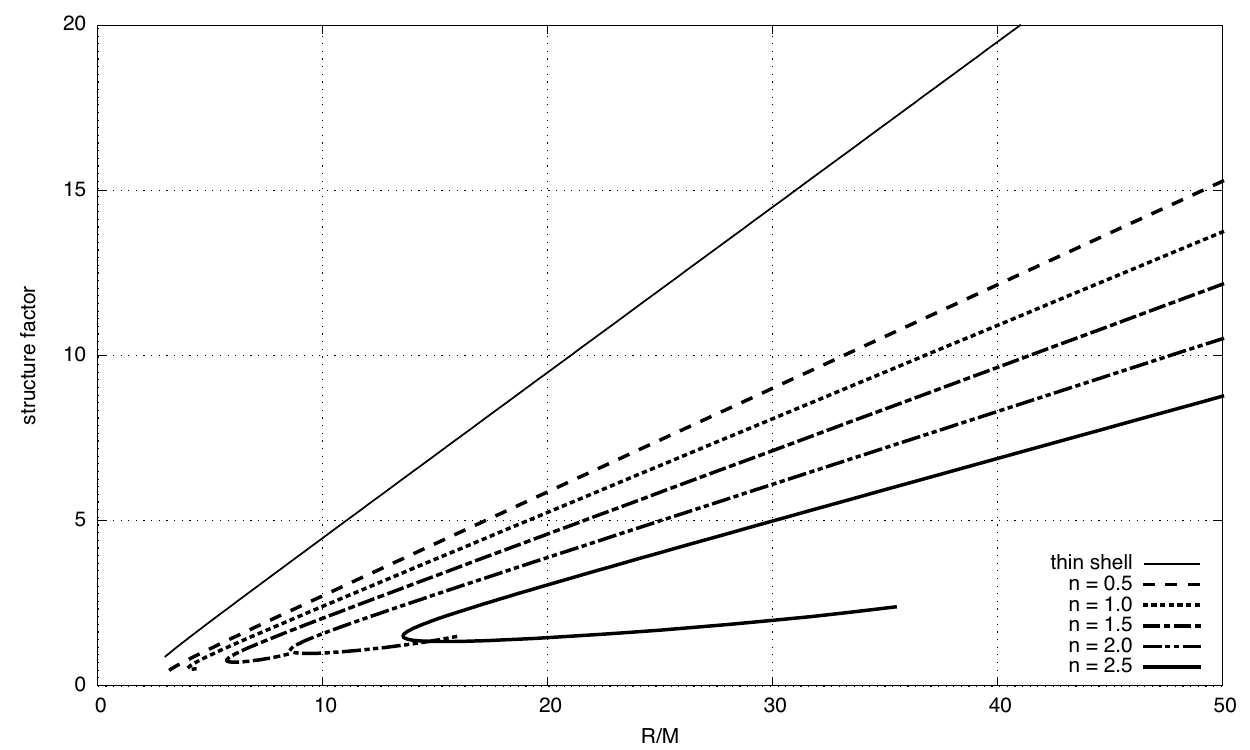}
\caption{Structure factor $S^{\rm scalar}_1$ for the
  scalar self-force, plotted as a function of $R/M$ for selected
  polytropic equations of state labelled by the polytropic index $n$.
  The curve for a massive thin shell is also displayed for
  comparison. At large $R/M$, the structure factor increases linearly
  with $R/M$, with a slope that depends on the polytropic index. For
  some equations of state the structure factor is multi-valued at
  small $R/M$; the low-lying branch, however, corresponds to stellar
  configurations beyond the maximum mass, which are dynamically
  unstable.}   
\label{fig:scalar_asymp1}
\end{figure}
 
When $r_0 \gg M$ the self-force differences are dominated by the
leading term $\ell = 1$ in the sums, and the expressions reduce to 
\begin{equation} 
\Delta F^r_{\rm scalar} \sim \frac{2}{3} S^{\rm scalar}_1 
\frac{q^2 M^3}{r_0^5}
\label{intro:scalar_asymp} 
\end{equation} 
and 
\begin{equation} 
\Delta F^r_{\rm em} \sim \frac{4}{3} S^{\rm em}_1 
\frac{e^2 M^3}{r_0^5}. 
\label{intro:em_asymp} 
\end{equation} 
This is the statement that to leading order in an expansion of the
self-force in powers of $r_0^{-1}$, the details of internal structure
are revealed at order $r_0^{-5}$; and these are contained in
the structure factors $S^{\rm scalar}_1$ and $S^{\rm em}_1$.    

In Fig.~\ref{fig:scalar_asymp1} we present plots of $S^{\rm scalar}_1$
as a function of $R/M$ for selected values of the polytropic index
$n$. We find that for large values of $R/M$, the structure factor
increases linearly with $R/M$, with a slope that depends on the 
polytropic index; these results imply that for fixed $r_0$ and $M$,
the self-force increases linearly with the stellar radius. This is the
same scaling that was found by Burko, Liu, and Soen
\cite{burko-etal:01} in the case of a massive thin shell, for which we
also plot the structure factor.  

\begin{figure}
\includegraphics[width=5in]{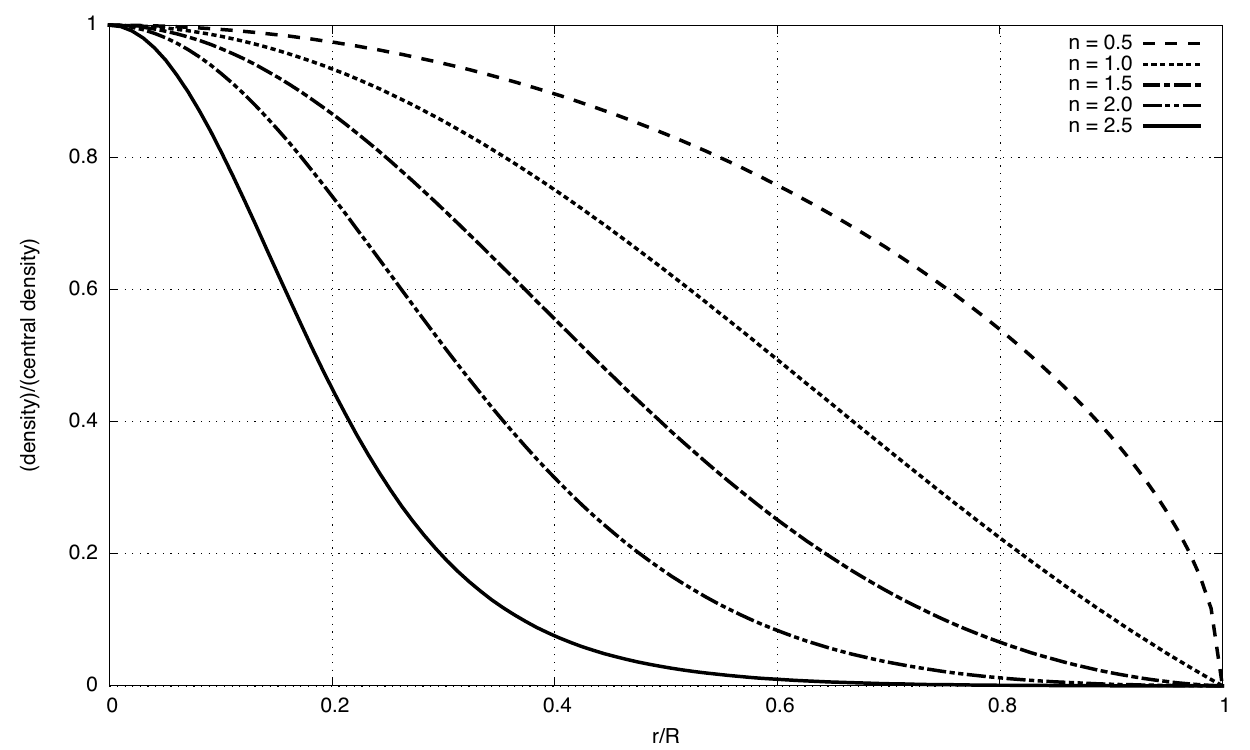}
\caption{Density profiles for selected polytropic models. The mass
  density $\rho$, normalized by the central density $\rho_c$, is
  plotted as a function of $r/R$. Each model corresponds to a body of
  radius-to-mass ratio $R/M \simeq 15$, and each curve is labelled
  by the polytropic index $n$. The corresponding values of the
  relativistic parameter $b$ (defined in Sec.~\ref{sec:polytrope}) are
  3.830543764152575e-2 for $n=0.5$, 4.116715459636999e-2 for $n=1.0$,
  4.733363787126200e-2 for $n=1.5$, 5.977719194440032e-2 for $n=2.0$, 
  and 9.929978563520857e-2 for $n=2.5$. The figure reveals that as $n$
  increases, the body becomes increasingly centrally dense.}  
\label{fig:density}
\end{figure}

An increase of the self-force with the stellar radius is to be
expected. The self-force results from an interation between the
field produced by the scalar charge and the curvature of
spacetime. Since the self-force would vanish in the pure Schwarzschild
spacetime of a black hole, the interaction is limited to the region of
spacetime occupied by the matter. For a fixed $M$, the region of
interaction increases with the stellar radius $R$, and it follows that
the self-force should increase with $R$. This argument, however, does
not explain why the enhancement with size is linear in $R$ instead of
some other relationship. The argument continues to apply to the case
of a massive thin shell, for which the curvature is entirely
concentrated within the shell; here also the size of the interaction
region increases with the shell's radius, leading again to an increase
of the self-force.   

\begin{figure}
\includegraphics[width=5in]{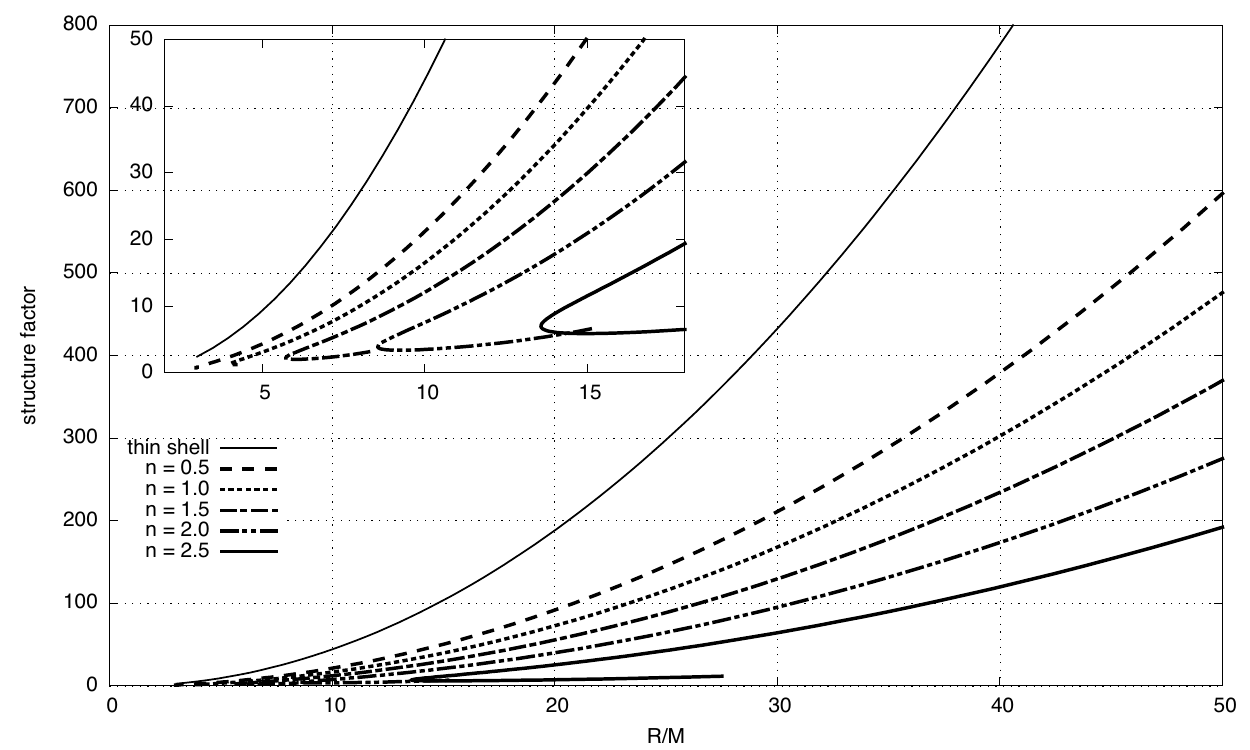}
\caption{Structure factor $S^{\rm em}_1$ for the 
  electromagnetic self-force, plotted as a function of $R/M$ for
  selected polytropic equations of state labelled by the polytropic
  index $n$. The curve for a massive thin shell is also displayed
  for comparison. At large $R/M$, the structure factor increases
  quadratically with $R/M$, with a curvature that depends on the
  equations of state.}   
\label{fig:em_asymp1}
\end{figure}
 
In addition to the linear scaling of the self-force with $R/M$, the
curves displayed in Fig.~\ref{fig:scalar_asymp1} reveal that for a fixed
$R/M$, the self-force is a decreasing function of the polytropic
index. An explanation for this effect can be deduced by extending the
preceding argument to take into account the distribution of mass
within the star. In Fig.~\ref{fig:density} we show density profiles
for polytropic models that share the same ratio $R/M$; one sees that
models with larger values of $n$ are more centrally dense than models
with smaller values. A centrally-dense star will possess a smaller
region of effective interation with the scalar field than a more
uniform star, and this will produce a smaller self-force.   

In Fig.~\ref{fig:em_asymp1} we present plots of $S^{\rm em}_1$ as a 
function of $R/M$ for selected equations of state. We find that for
large values of $R/M$, the structure factor increases quadratically with
$R/M$, with a curvature that depends on the polytropic index; these
results imply that for fixed $r_0$ and $M$, the self-force increases
quadratically with the stellar radius. This is the same scaling that 
was found by Burko, Liu, and Soen \cite{burko-etal:01} in the case of
a massive thin shell. 

The increase of the self-force difference with the stellar radius can
be explained in the same way as for the scalar case. Here the
self-force difference results from an interaction between the
electromagnetic field and the matter distribution, and a larger star
gives rise to a larger region of interaction, and therefore a larger
self-force difference; the argument does not explain why the
enhancement with size is quadratic in $R$ in the electromagnetic
case. The property that the self-force difference, for a fixed $R/M$,
is a decreasing function of the polytropic index, is also explained in
the same way, with the help of Fig.~\ref{fig:density}. Here also the 
property is a consequence of the fact that stellar models with larger
values of $n$ are more centrally dense than models with smaller
values. 

The main conclusion of this work is that aspects of the internal
structure of a spherical star can be revealed in a close examination
of the self-force at order $r_0^{-5}$. Our computations show that
the self-force increases with $R/M$, in a linear manner for the scalar
case, and in a quadratic manner for the electromagnetic case. In
addition, we find that the rate of increase depends on the equation of
state, a centrally dense body producing a smaller rate of increase
than a more uniform body. 

In the remainder of the paper we establish the results
presented previously. We begin in Sec.~\ref{sec:polytrope} with a
description of relativistic polytropes. In Sec.~\ref{sec:scalar} we
describe the integration of the scalar-field equation in the spacetime
of a massive body, and the computation of the self-force. In
Sec.~\ref{sec:em} we describe the calculations for the case of an
electric charge.  

\section{Polytropic stellar models} 
\label{sec:polytrope} 

In this section we describe the polytropic stellar models that are
involved in our examination of the influence of internal structure on
the scalar and electromagnetic self-force. Our models are static and
spherically symmetric, and we write the spacetime metric as  
\begin{equation} 
ds^2 = -e^{2\psi}\, dt^2 + f^{-1} dr^2 
+ r^2\bigl( d\theta^2 + \sin^2\theta\, d\phi^2 \bigr), 
\label{ds2_polytrope} 
\end{equation} 
with $\psi$ a function of $r$ and $f := 1-2m(r)/r$. For a perfect
fluid at rest the field equations are 
\begin{eqnarray} 
\label{poly_fieldeqns} 
\frac{dm}{dr} &=& 4\pi r^2 \sigma, \\ 
\frac{d\psi}{dr} &=& \frac{1}{r^2 f} \bigl( m + 4\pi r^3 p \bigr),\\ 
\frac{dp}{dr} &=& -(\sigma+p) \frac{d\psi}{dr} 
= -\frac{\sigma+p}{r^2 f} \bigl( m + 4\pi r^3 p \bigr), 
\end{eqnarray}
in which $\sigma$ is the energy density and $p$ the pressure; the
last equation is the condition of hydrostatic equilibrium. 

The polytropic equations of state are 
\begin{equation} 
p = K \rho^{1+1/n}, \qquad \epsilon = n p, 
\label{EOS} 
\end{equation} 
in which $K$ and $n$ are constants, $\rho$ is the rest-mass density,
and $\epsilon$ the internal thermodynamic energy; the total energy
density is then $\sigma = \rho + \epsilon$. 

To integrate the field equations we introduce the scaling quantities
$\rho_c$ (the central mass density), $p_c := K \rho_c^{1+1/n}$ (the
central pressure), $m_0 := 4\pi \rho_c r_0^3$ (a mass scale), and the
squared length scale $r_0^2 := (n+1) p_c/(4\pi \rho_c^2)$. We
introduce also the dimensionless relativistic parameter 
\begin{equation} 
b := \frac{p_c}{\rho_c} = K \rho_c^{1/n}, 
\label{b_def}
\end{equation} 
in terms of which $m_0/r_0 = (n+1) p_c/\rho_c = (n+1) b$. We make use
of the dimensionless variables $\xi$, $\theta$, and $\mu$ such that 
$r = r_0 \xi$, $\rho = \rho_c \theta^n$, and $m = m_0 \mu$. In terms
of the Lane-Emden variable $\theta$ we also have $p = p_c
\theta^{n+1}$, $\epsilon = n p_c \theta^{n+1}$, and $\sigma =
\rho_c (1+nb\theta) \theta^n$. 

The field equations become
\begin{eqnarray} 
\label{poly_LE1} 
\frac{d\mu}{d\xi} &=& \xi^2 (1 + n b \theta) \theta^n, \\ 
\frac{d\theta}{d\xi} &=& -\frac{1}{\xi^2 f} 
\bigl[ 1 + (n+1) b \theta \bigr] 
\bigl( \mu + b \xi^3 \theta^{n+1} \bigr),\\
\frac{d\psi}{d\xi} &=& \frac{(n+1) b}{\xi^2 f}   
\bigl( \mu + b \xi^3 \theta^{n+1} \bigr), 
\end{eqnarray} 
with $f = 1-2(n+1)b\mu/\xi$. The equations are integrated outward from
$\xi=0$, at which we impose the boundary conditions $\mu = 0$, 
$\theta = 1$, and $\psi = \psi_c$. Integration stops at $\xi = \xi_1$,
at which $\mu = \mu_1$, $\theta = 0$, and $\psi = \psi_1$. The star's 
radius is then $R = r_0 \xi_1$, its mass is $M = m_0 \mu_1$, and the
value of $\psi_c$ is chosen so that $e^{2\psi_1} = 1-2M/R$, to ensure
that the solution joins smoothly to the Schwarzschild metric when 
$r > R$.  

The numerical integration of the equations is facilitated by using
$\nu := \mu/\xi^3$ as a substitute for the mass function, and $x :=
\ln\xi$ as a substitute for the radial variable. In terms of these we
have 
\begin{eqnarray} 
\label{poly_LE2} 
\frac{d\nu}{dx} &=& (1+n b \theta) \theta^n - 3\nu, \\ 
\frac{d\theta}{dx} &=& -\frac{\xi^2}{f}   
\bigl[ 1 + (n+1) b \theta \bigr] 
\bigl( \nu + b \theta^{n+1} \bigr),\\
\frac{d\psi}{dx} &=& (n+1) b \frac{\xi^2}{f}   
\bigl( \nu + b \theta^{n+1} \bigr), 
\end{eqnarray} 
with $f = 1-2(n+1)b\xi^2\nu$ and $\xi = e^x$. The boundary
conditions are now placed at $x=-\infty$, at which $\nu =
\frac{1}{3}(1+nb)$, $\theta = 1$, and $\psi = \psi_c$. In practice the
integrations are started at $x = x_0 < 0$ such that $\xi_0 = e^{x_0}$
is very small, and starting values for $\nu$, $\theta$, and $\psi$ are 
obtained by expanding each quantity in powers of $\xi^2$ and
determining the coefficients with the help of the differential
equations. 

\begin{figure}
\includegraphics[width=5in]{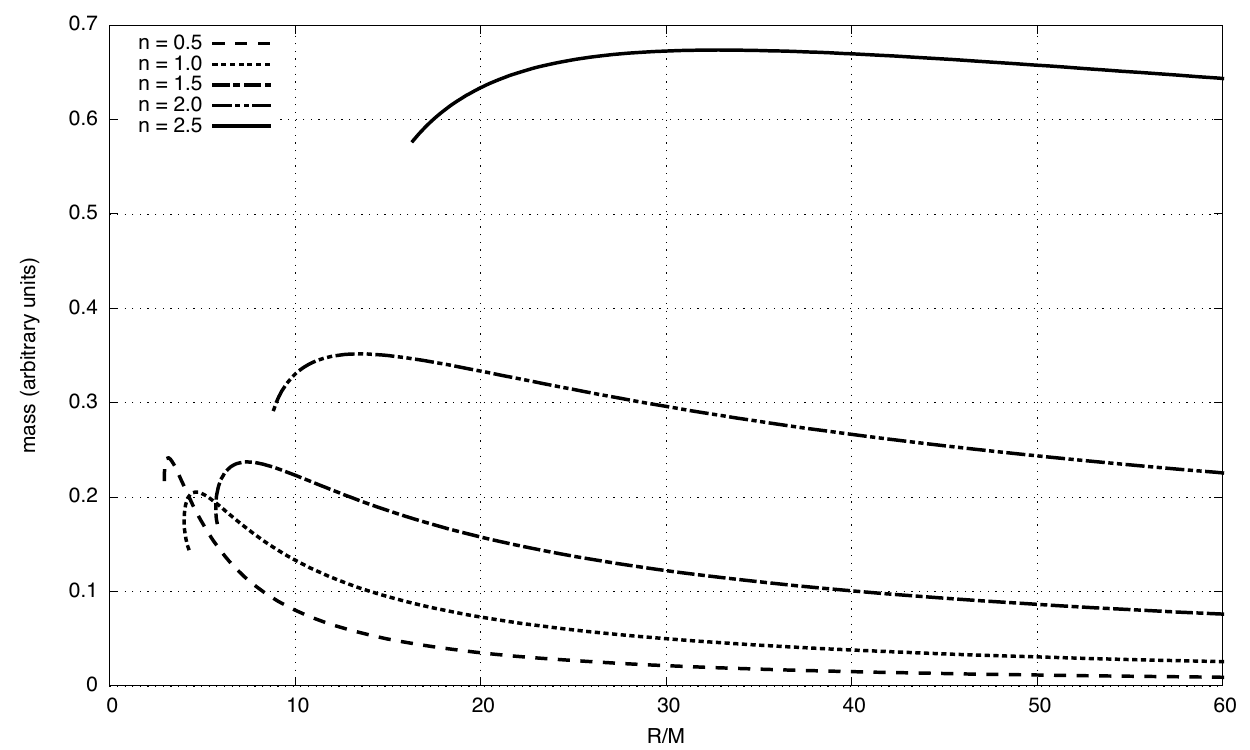}
\caption{Mass of a relativistic polytrope, in units of $\bar{m}_0$, as
  a function of $R/M$, for selected values of the polytropic index
  $n$. The lowest curve corresponds to $n=0.5$, higher curves
  correspond to increasing values of $n$, and the highest curve
  corresponds to $n = 2.5$. In each case the mass reaches a maximum
  value at a minimum value of $R/M$. Configurations with smaller $R/M$
  are dynamically unstable and therefore unphysical.}   
\label{fig:polytrope}
\end{figure}
 
Integration of the field equations for a selected value of the
polytropic index $n$ gives rise to a continuous sequence of stellar
models parameterized by $b$, which acts as a substitute for the
central density $\rho_c$. Because $m_0$ and $r_0$ depend on the
central density (and therefore on $b$), it is necessary to rescale the
mass and length units so as to eliminate this dependence before
producing plots of the mass $M$ and radius $R$ as functions of $b$ on 
the sequence. We therefore set 
\begin{equation} 
M = \bar{m}_0 b^{(3-n)/2} \mu_1, \qquad 
R = \bar{r}_0 b^{(1-n)/2} \xi_1, 
\end{equation} 
in which 
\begin{eqnarray}
\bar{m}_0 &:=& m_0 b^{-(3-n)/2} 
= \frac{ (n+1)^{3/2} K^{n/2} }{ (4\pi)^{1/2} }, \\ 
\bar{r}_0 &:=& r_0 b^{-(1-n)/2} 
= \frac{ (n+1)^{1/2} K^{n/2} }{ (4\pi)^{1/2} } 
\end{eqnarray}
are the rescaled mass and length units, respectively, which are
independent of $b$ and therefore constant on the sequence. Plots of
$M/\bar{m}_0$ as a function of $R/M$ for selected values of the
polytropic index $n$ are presented in Fig.~\ref{fig:polytrope}.    

\section{Scalar self-force} 
\label{sec:scalar} 

\subsection{Scalar field and self-force in spherical spacetimes}  

We consider the self-force acting on a static scalar charge $q$ moving
on a world line $c$ in a curved spacetime; the world line is
described by the parametric equations $x^\alpha = z^\alpha(\tau)$, in
which $\tau$ is proper time. The charge produces a scalar potential
$\Phi$ that satisfies the wave equation   
\begin{equation} 
\Box \Phi = -4\pi q \int_c \delta_4\bigl( x, z(\tau) \bigr)\,
d\tau, 
\label{scalar_wave} 
\end{equation} 
in which $\Box := g^{\alpha\beta} \nabla_\alpha \nabla_\beta$ is the
covariant wave operator, and $\delta_4(x,z)$ is a scalarized Dirac
distribution. The self-force acting on the scalar charge is given by 
\begin{equation} 
F^\alpha = q \bigl( g^{\alpha\beta} + u^\alpha u^\beta \bigr) 
\nabla_\beta \Phi^{\sf R}, 
\label{scalar_selfforce1} 
\end{equation} 
in which $u^\alpha := dz^\alpha/d\tau$ is the charge's velocity
vector, and $\Phi^{\sf R} := \Phi - \Phi^{\sf S}$ is the difference
between $\Phi$ and the Detweiler-Whiting singular potential 
$\Phi^{\sf S}$\cite{detweiler-whiting:03}; the regular potential
$\Phi^{\sf R}$ is known to be smooth on $c$, and to be
solely responsible for the self-force.   

For our particular application we take the spacetime to be static and
spherically symmetric, we write the metric as in
Eq.~(\ref{ds2_polytrope}), and we choose $c$ to be a static
world line at $r = r_0 > R$. We decompose the potential in spherical
harmonics,  
\begin{equation} 
\Phi(r,\theta,\phi) = \sum_{\ell m} \Phi_{\ell m}(r) 
Y_{\ell m}(\theta,\phi),  
\label{scalar_Ylm} 
\end{equation} 
and substitution within the wave equation produces 
\begin{equation} 
\fl
r^2 \frac{d^2 \Phi_{\ell 0}}{dr^2} 
+ \biggl( 2 + \frac{r}{2f} \frac{df}{dr}+ r \frac{d\psi}{dr} \biggr) 
r \frac{d\Phi_{\ell 0}}{dr} - \frac{\ell(\ell+1)}{f} \Phi_{\ell 0} 
= -4\pi q \sqrt{\frac{2\ell + 1}{4\pi}} f_0^{-1/2}\, \delta(r-r_0),  
\label{scalar_l0} 
\end{equation} 
in which $f_0 := f(r_0)$. Without loss of generality we placed the
charge on the axis $\theta = 0$, and exploited the property 
$Y_{\ell m}(0,\phi) = \sqrt{(2\ell+1)/(4\pi)} \delta_{m,0}$ of
spherical-harmonic functions. The modes $m \neq 0$ of the scalar
potential vanish.  

It is easy to show that the time and angular components of the
self-force vanish in a static and spherically-symmetric situation, and 
that the radial component is given by 
\begin{equation} 
F^r = q f_0 \lim_{x \to z} \sum_{\ell} 
\Bigl[ \bigl( \partial_r \Phi \bigr)_\ell 
- \bigl( \partial_r \Phi^{\sf S} \bigr)_\ell \Bigr], 
\label{scalar_selfforce2} 
\end{equation} 
in which 
\begin{equation} 
\bigl( \partial_r \Phi \bigr)_\ell := 
\sum_{m=-\ell}^\ell \frac{d\Phi_{\ell m}}{dr}(r) Y_{\ell m}(\theta,\phi) 
\end{equation} 
are the {\it multipole coefficients} of $\partial_r \Phi$, while
$( \partial_r \Phi^{\sf S} )_\ell$ are those of the
singular potential $\Phi^{\sf S}$. The limit in
Eq.~(\ref{scalar_selfforce2}) can be taken by setting  
$r = r_0 + \Delta$, $\theta = 0$, and letting $\Delta \to 0$ (from 
either direction). With this choice, we find that 
\begin{equation} 
\bigl( \partial_r \Phi \bigr)_\ell = 
\sqrt{\frac{2\ell+1}{4\pi}} \frac{d \Phi_{\ell 0}}{dr} (r_0 + \Delta).  
\end{equation} 
The multiplole coefficients can therefore be obtained from the
solution to Eq.~(\ref{scalar_l0}). This equation must be integrated in
the stellar interior of the polytropic model, and in the body's 
Schwarzschild exterior.   

\subsection{Interior} 

The equation satisfied by the $\Phi_{\ell 0}$ mode of the scalar
potential was given above in Eq.~(\ref{scalar_l0}), and since the
charge is situated in the exterior portion of the spacetime 
($r_0 > R$), the right-hand side of the equation is zero everywhere
within the interior. To integrate the equation we implement a change
of variables from $\Phi_{\ell 0}$ to $\eta_\ell$ defined by  
\begin{equation} 
\eta_\ell := \frac{d \ln\Phi_{\ell 0}}{d \ln r}. 
\label{scalar_etadef} 
\end{equation} 
The equation for $\Phi_{\ell 0}$ becomes 
\begin{equation} 
\xi \frac{d\eta_\ell}{d\xi} + \eta_\ell (\eta_\ell -1)  
+ {\cal D} \eta_\ell - \frac{\ell(\ell+1)}{f} = 0  
\label{scalar_etaeq}
\end{equation} 
when expressed in terms of $\eta_\ell$ and $\xi := r/r_0$, with 
${\cal D} := 2 + \frac{1}{2} f^{-1} r(df/dr) + r (d\psi/dr)$ given
explicitly by  
\begin{equation} 
{\cal D} = \frac{1}{f} \biggl\lgroup 2 - (n+1)b\xi^2 
\Bigl\{ 2\nu + \bigl[ 1 + (n-1)b\theta \bigr] \theta^n \Bigr\}
\biggr\rgroup.   
\end{equation}
The equation is integrated outward from $\xi = 0$, at which 
$\eta_\ell = \ell$. Here also the numerical treatment can be improved
by adopting $x = \ln\xi$ as a new independent variable, and adding a
few more terms to an expansion of $\eta_\ell$ in powers of
$\xi^2$. Integration proceeds until $\xi = \xi_1$, at which 
$\eta_\ell = \eta_\ell^1$. As we shall see, the number $\eta_\ell^1$
is the entirety of the information required about the internal
solution.   

\subsection{Exterior} 

Outside the matter the metric is described by the Schwarzschild
solution, so that $e^{2\psi} = f = 1-2M/r$. Equation (\ref{scalar_l0})
becomes 
\begin{equation} 
\fl
r^2 f \frac{d^2\Phi}{dr^2} 
+ 2\biggl(1-\frac{M}{r} \biggr) r \frac{d\Phi}{dr} - \ell(\ell+1) \Phi  
= -4\pi q \sqrt{\frac{2\ell+1}{4\pi}} f_0^{1/2} \delta(r-r_0), 
\label{scalarl0_ext} 
\end{equation} 
where we suppress the label ``$\ell 0$'' on $\Phi$. It is helpful to
use $z := r/M -1$ instead of $r$ as a radial variable, and 
to re-express the differential equation as 
\begin{equation}   
\fl
(z^2-1) \frac{d^2 \Phi}{dz^2} + 2z \frac{d\Phi}{dz} 
- \ell(\ell+1) \Phi  
= -4\pi \frac{q}{M} \sqrt{\frac{2\ell+1}{4\pi}} 
\biggl( \frac{z_0-1}{z_0+1} \biggr)^{1/2} 
\delta(z-z_0), 
\end{equation} 
where $z_0 := r_0/M - 1$. Away from $z_0$ the right-hand side of the
equation vanishes, and the solution is a linear superposition of
Legendre functions $P_\ell(z)$ and $Q_\ell(z)$. The presence of the
delta function implies that the solution must respect the junction
conditions 
\begin{equation} 
\fl
\bigl[ \Phi \bigr] = 0, \qquad 
\biggl[ \frac{d\Phi}{dz} \biggr] = -4\pi \frac{q}{M}
\sqrt{\frac{2\ell+1}{4\pi}}  
\frac{1}{(z_0-1)^{1/2} (z_0+1)^{3/2}} 
\label{scalar_jumps} 
\end{equation} 
at $z = z_0$; here $[f] := f(z=z_0 + \epsilon) - f(z=z_0 - \epsilon)$,
with $\epsilon$ approaching zero from above, is the jump of the
function $f$ across $z=z_0$. 

\subsection{Matching} 

The interior and exterior solutions must match smoothly at $r = R$, or
$z = Z := R/M - 1$, and the two exterior solutions (for $z < z_0$ and
$z > z_0$) must satisfy the junction conditions at $z = z_0$. For 
$z < Z$ we have that $\Phi = \Phi_{\rm in}$, for $Z < z < z_0$ we have
that $\Phi = A P_\ell(z) + B Q_\ell(z)$, where $A$ and $B$ are
coefficients to be determined, and for $z > z_0$ we have that
$\Phi = C Q_\ell(z)$, where $C$ is a third unknown coefficient; we
exclude a term in $P_\ell(z)$ when $z > z_0$ because it would produce
a diverging field at $z=\infty$. Letting $\alpha := A P_\ell(z_0)$, 
$\beta := B Q_\ell(z_0)$, $\gamma := C Q_\ell(z_0)$, and 
$\delta := \Phi_{\rm in}(Z)$, we find that the matching conditions 
give rise to the relations 
\begin{eqnarray} 
\alpha &=& (z_0-1)(z_0+1) P_\ell(z_0) Q_\ell(z_0)\, J, \\ 
\beta &=& -(z_0-1)(z_0+1) \frac{ (Z+1) P'_\ell(Z) - \eta_\ell^1
  P_\ell(Z) }{ (Z+1) Q'_\ell(Z) - \eta_\ell^1 Q_\ell(Z) }
\bigl[ Q_\ell(z_0)\bigr]^2 \, J, 
\end{eqnarray} 
in which 
\begin{equation} 
J := 4\pi \frac{q}{M}\sqrt{\frac{2\ell+1}{4\pi}} 
\frac{1}{(z_0-1)^{1/2} (z_0+1)^{3/2}}, 
\end{equation} 
along with $\gamma = \alpha + \beta$ and $\delta =
[P_\ell(Z)/P_\ell(z_0)] \alpha + [Q_\ell(Z)/Q_\ell(z_0)] \beta$. To
arrive at these results we used the Wronskian property of the Legendre
functions, $P_\ell Q'_\ell - Q_\ell P'_\ell = -(z^2-1)^{-1}$, with a
prime indicating differentiation with respect to $z$.  As we shall
see, the quantity of prime interest for a self-force computation is
$\gamma$.    

\subsection{Black-hole problem}  

If the scalar charge were placed outside a black hole instead of in
the exterior region of a spherical star, the Schwarzschild metric would 
apply everywhere instead of being restricted to $r > R$ or $z > Z$. In
such a case there would be no interior solution to the scalar-field
equation, and the two external solutions would be given by 
$\Phi = \bar{A} P_\ell(z)$ for $z < z_0$, and $\Phi = \bar{C}
Q_\ell(z)$ for $z > z_0$; a term in $Q_\ell(z)$ is excluded when 
$z < z_0$ because it would diverge on the event horizon. Letting
$\bar{\alpha} := \bar{A} P_\ell(z_0)$ and $\bar{\gamma} := \bar{C}
Q_\ell(z_0)$, we find that the junction conditions at $z=z_0$ produce 
$\bar{\alpha} = \bar{\gamma} = \alpha$.   
    
\subsection{Self-force difference} 

The radial component of the self-force acting on a scalar charge $q$
at $r = r_0$ is given by Eq.~(\ref{scalar_selfforce2}), which we
express as   
\begin{equation} 
F^r_{\rm star} = q f_0 \sum_{\ell}  \Bigl[ 
\bigl( \partial_r \Phi_{\rm star} \bigr)_\ell 
- \bigl( \partial_r \Phi^{\sf S} \bigr)_\ell \Bigr], 
\label{scalarf_star} 
\end{equation} 
with the understanding that the right-hand side is evaluated in the
limit $\Delta \to 0$. 

In addition to this self-force, we may also consider the self-force
acting on another scalar charge $q$ placed at $r = r_0$ in the pure
Schwarzschild spacetime of a black hole. For this situation we would
have instead  
\begin{equation} 
F^r_{\rm hole} = q f_0  \sum_{\ell}  \Bigl[ 
\bigl( \partial_r \Phi_{\rm hole} \bigr)_\ell 
- \bigl( \partial_r \Phi^{\sf S} \bigr)_\ell \Bigr],  
\label{scalarf_hole} 
\end{equation} 
which refers to $\Phi_{\rm hole}$, the scalar potential as computed in
the pure Schwarzschild spacetime, and {\it to the same singular
  potential} as in Eq.~(\ref{scalarf_star}). The singular potential is
the same in each case, because it depends on the structure of
spacetime only in the immediate vicinity of the charge, where it is
described by the Schwarzschild metric.  

Taking the difference between Eqs.~(\ref{scalarf_star}) and
(\ref{scalarf_hole}), we have that      
\begin{equation} 
\Delta F^r := F^r_{\rm star} - F^r_{\rm hole}
= q f_0  \sum_{\ell}  \Bigl[ 
\bigl( \partial_r \Phi_{\rm star} \bigr)_\ell 
- \bigl( \partial_r \Phi_{\rm hole} \bigr)_\ell \Bigr],  
\label{scalar_forcediff1}  
\end{equation} 
and we see that the {\it difference} between the two self-forces can
be computed without the involvement of the singular field; this method
of regularization was originally devised by Drivas and Gralla
\cite{drivas-gralla:11}. Because the black-hole force actually
vanishes \cite{wiseman:00}, we have that  
\begin{equation} 
F^r_{\rm star} = \Delta F^r,
\label{scalar_sf} 
\end{equation} 
and $\Phi_{\rm hole}$ can in fact be identified with the
Detweiler-Whiting singular potential.  

The force difference can now be expressed as 
\begin{equation} 
\Delta F^r = q f_0 \sum_{\ell = 0}^\infty 
\sqrt{\frac{2\ell+1}{4\pi}}\biggl[ 
\frac{d\Phi_{\rm star}}{dr}(r_0 + \Delta)  
- \frac{d\Phi_{\rm hole}}{dr}(r_0 + \Delta) \biggr], 
\end{equation} 
in which each $\Phi$ stands for the $\ell 0$ mode of the
spherical-harmonic expansion. While each individual mode-sum 
for $F^r_{\rm star}$ and $F^r_{\rm hole}$ would fail to converge, the
mode-sum for $\Delta F^r$ can be shown to converge exponentially.    

The mode-sum requires the evaluation of $d\Phi_{\rm star}/dr$ and  
$d\Phi_{\rm hole}/dr$ at $r =r_0 + \Delta$. It is convenient to take
$\Delta = 0^+$, so that $r = r_0$ is approached from above. We
therefore insert $Md\Phi_{\rm star}/dr = \gamma
Q'_\ell(z_0)/Q_\ell(z_0)$ and $M d\Phi_{\rm hole}/dr = \bar{\gamma}    
Q'_\ell(z_0)/Q_\ell(z_0)$, and obtain 
\begin{equation}  
\Delta F^r = \frac{q}{M} f_0 \sum_{\ell = 1}^\infty 
\sqrt{\frac{2\ell+1}{4\pi}} \bigl( \gamma - \bar{\gamma} \bigr) 
\frac{Q'_\ell(z_0)}{Q_\ell(z_0)}; 
\end{equation} 
the sum now excludes $\ell = 0$, because the monopole
solutions are the same in the two spacetimes and therefore do not
contribute to the difference. (The monopole solution describes the
field of a spherical shell of charge $q$ at $r=r_0$; in both
spacetimes the field vanishes inside the shell, and goes as $q/r^2$
outside the shell.) Inserting the result obtained previously for 
$\gamma - \bar{\gamma} = \beta$, we arrive at the expression of
Eq.~(\ref{intro:scalar_force}),  
\begin{equation} 
\Delta F_{\rm star} = -\Bigl( \frac{q}{M} \Bigr)^2 
\biggl( \frac{z_0-1}{z_0+1} \biggr)^{3/2} \sum_{\ell=1}^\infty 
(2\ell + 1) S_\ell Q_\ell(z_0) Q'_\ell(z_0), 
\label{scalar_force} 
\end{equation} 
in which 
\begin{equation} 
S_\ell := \frac{ (Z+1) P'_\ell(Z) - \eta_\ell^1
  P_\ell(Z) }{ (Z+1) Q'_\ell(Z) - \eta_\ell^1 Q_\ell(Z) }
\label{scalar_structure} 
\end{equation} 
is a structure factor that depends on the stellar model (through the
stellar radius $R$ and the interior constant $\eta_\ell^1$) but is
independent of $r_0$. We recall that $z := r/M - 1$, $Z := R/M -1$,
$z_0 := r_0/M - 1$, and that a prime indicates differentiation with
respect to $z$. The mode-sum can be evaluated straightforwardly once
the numerical results for $\eta_\ell^1$ and $R/M$ are available.   

\begin{figure}
\includegraphics[width=5in]{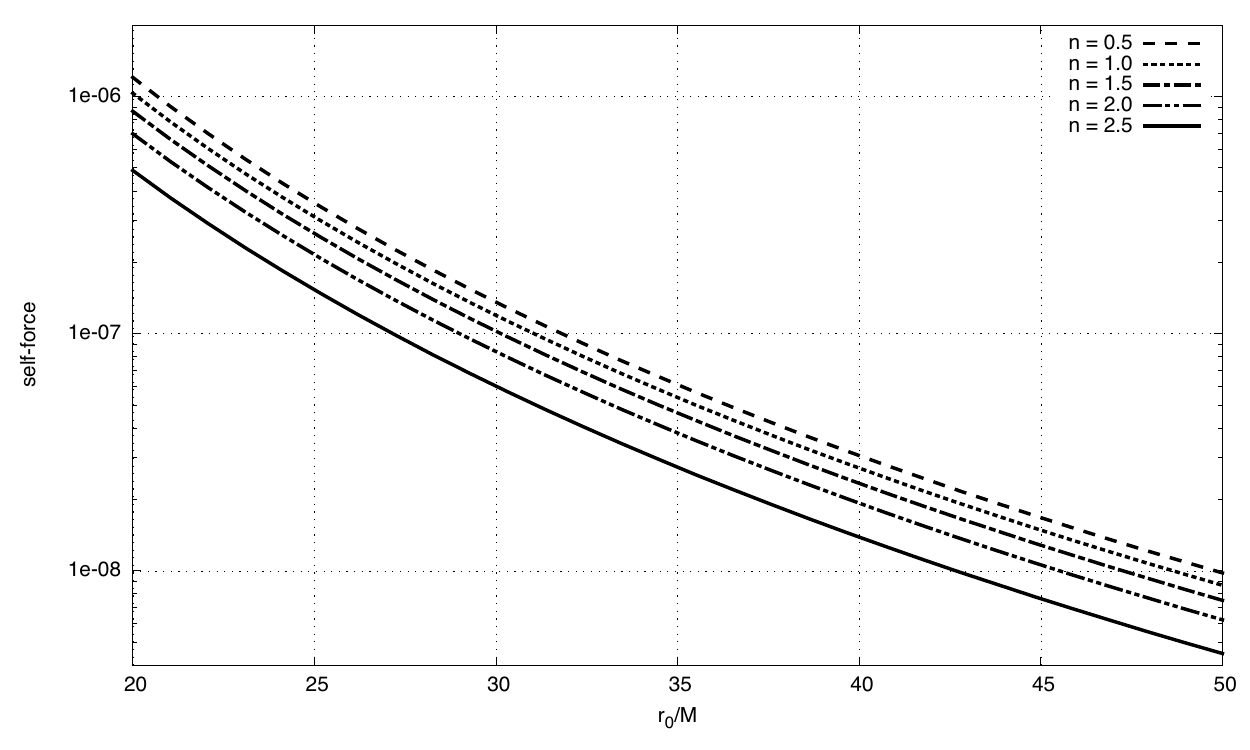}
\caption{Scalar self-force difference $\Delta F^r$, in units of
  $(q/M)^2$, plotted as a function of $r_0/M$ for selected polytropic
  equations of state labelled by the polytropic index $n$. Each
  stellar model corresponds to a body of radius-to-mass ratio 
  $R/M \simeq 15$ (see Fig~\ref{fig:density}). The falloff with $r_0$
  is consistent with the asymptotic behaviour $\sim r_0^{-5}$
  predicted by Eq.~(\ref{intro:scalar_asymp}), and for fixed $r_0/M$
  the self-force difference is seen to decrease with increasing $n$;
  this behaviour is discussed and explained in Sec.~\ref{sec:intro}.}  
\label{fig:scalar_force}
\end{figure}

The self-force of Eq.~(\ref{scalar_force}) can be computed as a 
function of $r_0$ for selected stellar models characterized by a 
polytropic index $n$ and a relativistic parameter $b$. Sample results
are displayed in Fig.~\ref{fig:scalar_force}, and our findings were
presented more fully in Sec.~\ref{sec:intro}. The self-force for a
massive thin shell can be computed by integrating the scalar-field
equation in the flat interior and matching the solution to the
Schwarzschild exterior; this yields 
$\eta_\ell^1 = \ell/\sqrt{1-2M/R}$, which can be 
inserted within Eq.~(\ref{scalar_structure}) and shown to reproduce
the results of Burko, Liu, and Soen \cite{burko-etal:01}. 

\section{Electromagnetic self-force} 
\label{sec:em} 

\subsection{Electromagnetic field and self-force in spherical
  spacetimes}  

We next consider the self-force acting on a static electric charge $e$
moving on a world line $c$ in a curved spacetime. The charge
produces a vector potential $\Phi_\alpha$ that satisfies the wave
equation    
\begin{equation} 
\Box \Phi_{\alpha} - R_{\alpha}^{\ \beta} \Phi_\beta 
= -4\pi e \int_c u^\alpha \delta_4( x, z)\, d\tau 
\label{em_wave} 
\end{equation} 
in the Lorenz gauge $\nabla^\alpha \Phi_\alpha = 0$; here
$R_{\alpha\beta}$ is the spacetime's Ricci tensor. The potential gives
rise to the electromagnetic field $F_{\alpha\beta} = \nabla_\alpha
\Phi_\beta - \nabla_\beta \Phi_\alpha$, and the self-force acting on
the electric charge is given by   
\begin{equation} 
F^\alpha = e F^{\ \alpha}_{{\sf R}\ \beta} u^\beta, 
\label{em_selfforce1} 
\end{equation} 
in which $F^{\ \alpha}_{{\sf R}\ \beta} := F^{\alpha}_{\ \beta}  
- F^{\ \alpha}_{{\sf S}\ \beta}$ is the difference between
the actual electromagnetic field and the Detweiler-Whiting singular
field \cite{detweiler-whiting:03}; the regular field 
$F^{\ \alpha}_{{\sf R}\ \beta}$ is known to be smooth on $c$, and 
to be solely responsible for the self-force.    

For our particular application we take the spacetime to be static and
spherically symmetric, we write the metric as in
Eq.~(\ref{ds2_polytrope}), and we choose $c$ to be a static
world line at $r = r_0 > R$. The only relevant component of the vector
potential is $\Phi_t$, and we decompose it as
\begin{equation} 
\Phi_t(r,\theta,\phi) = \sum_{\ell m} \Phi_{t\, \ell m}(r) 
Y_{\ell m}(\theta,\phi).  
\label{em_Ylm} 
\end{equation} 
Substitution within the wave equation produces 
\begin{eqnarray} 
&
r^2 \frac{d^2 \Phi_{t\, \ell 0}}{dr^2} 
+ \biggl( 2 + \frac{r}{2f} \frac{df}{dr} - r \frac{d\psi}{dr} \biggr) 
r \frac{d\Phi_{t\,\ell 0}}{dr} - \frac{\ell(\ell+1)}{f} \Phi_{t\, \ell 0} 
\nonumber \\ 
& \qquad 
= 4\pi e \sqrt{\frac{2\ell + 1}{4\pi}} e^{\psi_0} f_0^{-1/2}\,
\delta(r-r_0),   
\label{em_l0} 
\end{eqnarray} 
in which $f_0 := f(r_0)$ and $\psi_0 = \psi(r_0)$. Once more we placed
the  charge on the axis $\theta = 0$, and exploited the property 
$Y_{\ell m}(0,\phi) = \sqrt{(2\ell+1)/(4\pi)} \delta_{m,0}$ of
spherical-harmonic functions. The modes $m \neq 0$ of the potential
vanish.   

Inserting the decomposition of Eq.~(\ref{em_Ylm}) within
Eq.~(\ref{em_selfforce1}) yields  
\begin{equation} 
F^r = e e^{-\psi_0} f_0 \lim_{x \to z} \sum_{\ell} 
\Bigl[ \bigl( \partial_r \Phi_t \bigr)_\ell 
- \bigl( \partial_r \Phi^{\sf S}_t \bigr)_\ell \Bigr], 
\label{em_selfforce2} 
\end{equation} 
in which 
\begin{equation} 
\fl
\bigl( \partial_r \Phi_t \bigr)_\ell := 
\sum_{m=-\ell}^\ell \frac{d\Phi_{t\, \ell m}}{dr}(r) 
Y_{\ell m}(\theta,\phi)  
= \sqrt{\frac{2\ell+1}{4\pi}} 
\frac{d \Phi_{t\, \ell 0}}{dr} (r_0 + \Delta) 
\end{equation} 
are the {\it multipole coefficients} of $\partial_r \Phi_t$, while
$( \partial_r \Phi^{\sf S}_t )_\ell$ are those of the
singular potential $\Phi^{\sf S}_t$. The limit $x \to z$ is taken by
setting $r = r_0 + \Delta$, $\theta = 0$, and letting $\Delta \to 0$. 
The multiplole coefficients can be obtained from the solution to
Eq.~(\ref{em_l0}). As in the scalar case, this equation must be
integrated in the stellar interior of the polytropic model, and in the
body's Schwarzschild exterior.   

\subsection{Interior} 

The equation satisfied by the $\Phi_{t\, \ell 0}$ mode of the
electromagnetic potential was given above in Eq.~(\ref{em_l0}), and
since the charge is situated in the exterior portion of the spacetime
($r_0 > R$), the right-hand side of the equation is zero everywhere
within the interior. As in the scalar case we implement a change of
variables from $\Phi_{t\, \ell 0}$ to $\eta_\ell$ defined by  
\begin{equation} 
\eta_\ell := \frac{d \ln\Phi_{t\, \ell 0}}{d \ln r}.  
\label{em_etadef} 
\end{equation} 
The equation for $\Phi_{t\, \ell 0}$ becomes 
\begin{equation} 
\xi \frac{d\eta_\ell}{d\xi} + \eta_\ell (\eta_\ell -1)  
+ {\cal D} \eta_\ell - \frac{\ell(\ell+1)}{f} = 0, 
\label{em_etaeq}
\end{equation} 
with ${\cal D} := 2 + \frac{1}{2} f^{-1} r (df/dr) - r (d\psi/dr)$
given explicitly by  
\begin{equation} 
{\cal D} = \frac{1}{f} \biggl\lgroup 2 - (n+1)b\xi^2 
\Bigl\{ 4\nu + \bigl[ 1 + (n+1)b\theta \bigr] \theta^n \Bigr\}
\biggr\rgroup.   
\end{equation}
The equation is integrated outward from $\xi = 0$, at which 
$\eta_\ell = \ell$. Here also the numerical treatment is improved
by adopting $x = \ln\xi$ as new radial variable, and adding a
few more terms to an expansion of $\eta_\ell$ in powers of
$\xi^2$. Integration proceeds until $\xi = \xi_1$, at which 
$\eta_\ell = \eta_\ell^1$. 

\subsection{Exterior} 

Outside the matter the metric is described by the Schwarzschild
solution, so that $e^{2\psi} = f = 1-2M/r$. Equation (\ref{em_l0})
becomes 
\begin{equation} 
\fl
r^2 f \frac{d^2\Phi}{dr^2} 
+ 2r f\frac{d\Phi}{dr} - \ell(\ell+1) \Phi  
= 4\pi (e e^{\psi_0}) \sqrt{\frac{2\ell+1}{4\pi}} f_0^{1/2}
\delta(r-r_0),  
\end{equation} 
where we suppress the label ``$t\, \ell 0$'' on $\Phi$. The factor
$e^{\psi_0}$ is equal to $f_0^{1/2}$, but we prefer to lump it with
the electric charge $e$ in order to keep the right-hand side in the
same form as the right-hand side of Eq.~(\ref{scalarl0_ext}); in this
way the electromagnetic source is obtained directly from the scalar
source by the replacement $q \to -e e^{\psi_0}$, with $e e^{\psi_0}$
playing the role of an effective charge.  

Once again it is helpful to use $z := r/M -1$ instead of $r$ as a
radial variable, and to re-express the differential equation as  
\begin{equation}   
\fl
(z^2-1) \frac{d^2\Phi}{dz^2} + 2(z-1) \frac{d\Phi}{dz} 
- \ell(\ell+1) \Phi = 4\pi \frac{e e^{\psi_0}}{M} 
\sqrt{\frac{2\ell+1}{4\pi}} 
\biggl( \frac{z_0-1}{z_0+1} \biggr)^{1/2} \delta(z-z_0), 
\end{equation} 
where $z_0 := r_0/M - 1$. Away from $z_0$ the right-hand side of the
equation vanishes, and the solution is a linear superposition of
$(z-1) P'_\ell(z)$ and $(z-1) Q'_\ell(z)$, in a which a prime indicates
differentiation with respect to $z$. The presence of the
delta function implies that the solution must respect the junction
conditions 
\begin{equation}
\fl 
\bigl[ \Phi \bigr] = 0, \qquad 
\biggl[ \frac{d\Phi}{dz}\biggr] = 4\pi \frac{e e^{\psi_0}}{M} 
\sqrt{\frac{2\ell+1}{4\pi}} 
\frac{1}{(z_0-1)^{1/2} (z_0+1)^{3/2}} 
\label{em_jumps} 
\end{equation} 
at $z = z_0$; we recall that $[f] := f(z=z_0 + \epsilon) - f(z=z_0 -
\epsilon)$, is the jump of the function $f$ across $z=z_0$. 

\subsection{Matching} 

The interior and exterior solutions must match smoothly at $z = Z :=
R/M - 1$, and the two exterior solutions must satisfy the junction
conditions at $z = z_0$. For $z < Z$ we have that $\Phi = \Phi_{\rm
  in}$, for $Z < z < z_0$ we have that $\Phi = A (z-1) P'_\ell(z) 
+ B (z-1) Q'_\ell(z)$, where $A$ and $B$ are coefficients to be
determined, and for $z > z_0$ we have that $\Phi = C (z-1)
Q'_\ell(z)$, where $C$ is a third unknown coefficient. Letting 
$\alpha := A (z_0-1) P'_\ell(z_0)$, 
$\beta := B (z_0-1) Q'_\ell(z_0)$, 
$\gamma := C (z_0-1) Q'_\ell(z_0)$, and 
$\delta := \Phi_{\rm in}(Z)$, we find that the matching conditions
take the form of  
\begin{eqnarray} 
\fl
\alpha &=& -\frac{(z_0-1)^2(z_0+1)}{\ell(\ell+1)} P'_\ell(z_0)
Q'_\ell(z_0)\, J, \\  
\fl
\beta &=& \frac{(z_0-1)^2(z_0+1)}{\ell(\ell+1)} 
\frac{ \ell(\ell+1) P_\ell(Z) - (1+\eta^1_\ell)(Z-1) P'_\ell(Z) }
  { \ell(\ell+1) Q_\ell(Z) - (1+\eta^1_\ell)(Z-1) Q'_\ell(Z) } 
\bigl[ Q'_\ell(z_0)\bigr]^2 \, J, 
\end{eqnarray} 
in which 
\begin{equation} 
J := -4\pi \frac{e e^{\psi_0}}{M}\sqrt{\frac{2\ell+1}{4\pi}} 
\frac{1}{(z_0-1)^{1/2} (z_0+1)^{1/2}}, 
\end{equation} 
along with $\gamma = \alpha + \beta$ and $\delta =
\{[(Z-1)P'_\ell(Z)]/[(z_0-1)P'_\ell(z_0)]\} \alpha 
+ \{[(Z-1)Q'_\ell(Z)]/[(z_0-1)Q'_\ell(z_0)]\} \beta$. As in the scalar
case, the quantity of most direct interest is $\gamma$.  

\subsection{Black-hole problem}  

If the electric charge were placed outside a black hole instead of in 
the vacuum region of a spherical star, the Schwarzschild metric would 
apply everywhere instead of being restricted to $r > R$ or $z > Z$. In
such a case the two external solutions would be given by 
$\Phi = \bar{A} (z-1) P'_\ell(z)$ for $z < z_0$, and $\Phi = \bar{C}
(z-1) Q'_\ell(z)$ for $z > z_0$. Letting 
$\bar{\alpha} := \bar{A} (z_0-1) P'_\ell(z_0)$ and 
$\bar{\gamma} := \bar{C} (z_0-1) Q'_\ell(z_0)$, we find that the
junction conditions at $z=z_0$ produce $\bar{\alpha} = \bar{\gamma} 
= \alpha$.  
    
\subsection{Self-force difference} 

As in the scalar case we consider the difference between two
self-forces,   
\begin{equation} 
\Delta F^r := F^r_{\rm star} - F^r_{\rm hole}
= e e^{-\psi_0} f_0  \sum_{\ell}  \Bigl[ 
\bigl( \partial_r \Phi_{\rm star} \bigr)_\ell 
- \bigl( \partial_r \Phi_{\rm hole} \bigr)_\ell \Bigr],  
\label{em_forcediff1}  
\end{equation} 
the first acting on a charge $e$ at position $r_0$ in the exterior of
a material body, and the second acting on an identical charge at the
same position outside a Schwarzschild black hole. Because the
black-hole force is given by the Smith-Will expression $F^r_{\rm hole}
= e^2 M f_0^{1/2}/r_0^3$ \cite{smith-will:80}, we have that the
stellar force is given by    
\begin{equation} 
F^r_{\rm star} = \frac{e^2 M}{r_0^3} f_0^{1/2} + \Delta F^r. 
\label{em_sf} 
\end{equation} 
Apart from a monopole piece that is entirely responsible for the
black-hole force, $\Phi_{\rm hole}$ can be identified with the
Detweiler-Whiting singular potential. 

The force difference can be expressed as 
\begin{equation} 
\fl
\Delta F^r = e e^{-\psi_0} f_0 \sum_{\ell = 0}^\infty 
\sqrt{\frac{2\ell+1}{4\pi}}\Bigl[ 
\frac{d\Phi_{\rm star}}{dr}(r_0 + \Delta)  
- \frac{d\Phi_{\rm hole}}{dr}(r_0 + \Delta) \Bigr],  
\end{equation} 
in which each $\Phi$ stands for the $\ell 0$ mode of the
spherical-harmonic expansion of the vector potential $\Phi_t$. 
The mode-sum requires the evaluation of $d\Phi_{\rm star}/dr$ and   
$d\Phi_{\rm hole}/dr$ at $r =r_0 + \Delta$. It is convenient to take
$\Delta = 0^+$, so that $r = r_0$ is approached from above. We insert
the appropriate expressions within $\Delta F^r$ and obtain  
\begin{equation} 
\fl
\Delta F^r = \frac{e e^{-\psi_0}}{M} \frac{f_0}{z_0-1} 
\sum_{\ell = 1}^\infty \sqrt{\frac{2\ell+1}{4\pi}} 
\bigl( \gamma - \bar{\gamma} \bigr) 
\biggl[ -1 + \frac{\ell(\ell+1)}{z_0-1} 
  \frac{Q_\ell(z_0)}{Q'_\ell(z_0)} \biggr],    
\end{equation} 
in which a prime indicates differentiation with respect to 
$z := r/M - 1$; the sum now excludes $\ell = 0$, because the monopole
solutions are the same in the two spacetimes and therefore do not
contribute to the difference. Inserting the result obtained previously
for $\gamma - \bar{\gamma} = \beta$, we arrive at the expression of
Eq.~(\ref{intro:em_force}),   
\begin{equation}
\fl 
\Delta F^r = -\Bigl( \frac{e}{M} \Bigr)^2 
\biggl( \frac{z_0-1}{z_0+1} \biggr)^{3/2} \sum_{\ell=1}^\infty 
(2\ell + 1) S_\ell \biggl[ Q_\ell(z_0) 
- \frac{(z_0-1)Q'_\ell(z_0)}{\ell(\ell+1)} \biggr] 
Q'_\ell(z_0), 
\label{em_force} 
\end{equation} 
in which 
\begin{equation} 
S_\ell := 
\frac{ \ell(\ell+1) P_\ell(Z)  - (1+\eta^1_\ell)(Z-1) P'_\ell(Z) } 
  { \ell(\ell+1) Q_\ell(Z) - (1+\eta^1_\ell)(Z-1) Q'_\ell(Z) } 
\label{em_structure} 
\end{equation} 
is a structure factor that depends on the stellar model (through the
stellar radius $R$ and the interior constant $\eta_\ell^1$) but is
independent of $r_0$. We recall that $z := r/M - 1$, $Z := R/M -1$,
$z_0 := r_0/M - 1$, and that a prime indicates differentiation with
respect to $z$. The mode-sum can be evaluated straightforwardly once
the numerical results for $\eta_\ell^1$ and $R/M$ are available.   

\begin{figure}
\includegraphics[width=5in]{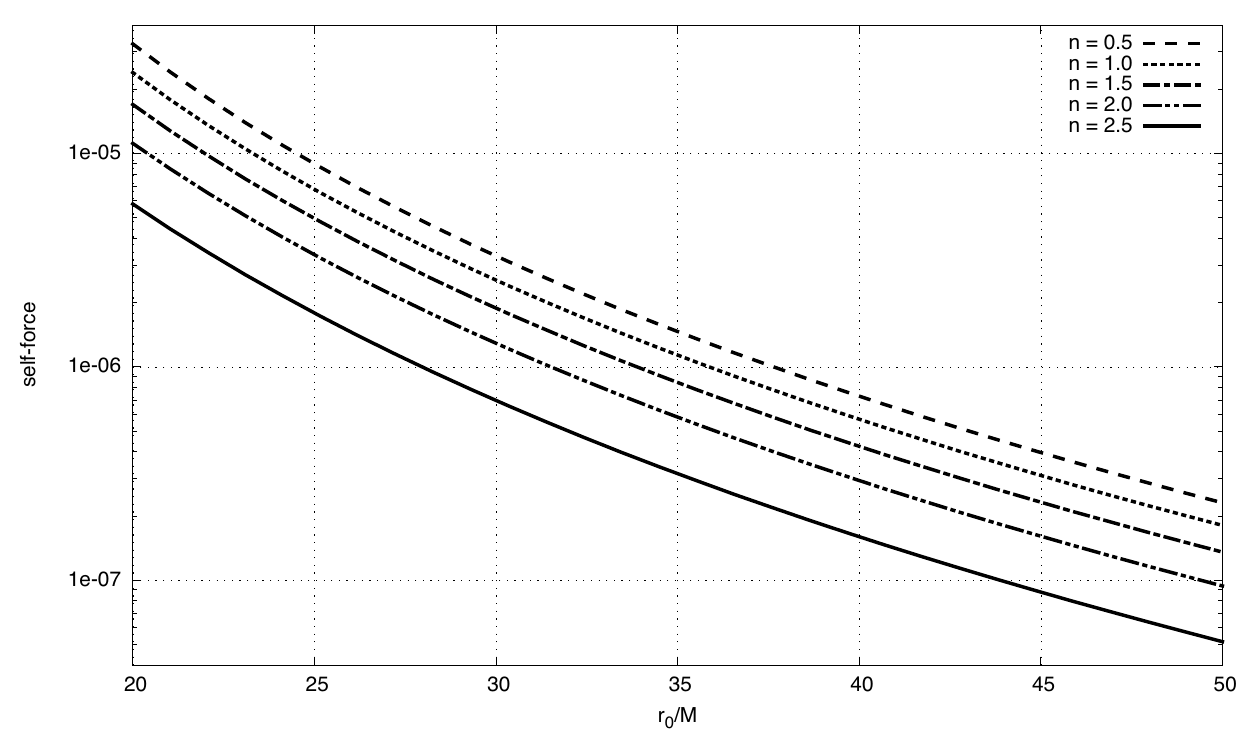}
\caption{Electromagnetic self-force difference $\Delta F^r$, in units
  of $(e/M)^2$, plotted as a function of $r_0/M$ for selected
  polytropic equations of state labelled by the polytropic index
  $n$. Each stellar model corresponds to a body of radius-to-mass
  ratio $R/M \simeq 15$ (see Fig~\ref{fig:density}). The falloff with
  $r_0$ is consistent with the asymptotic behaviour $\sim r_0^{-5}$ 
  predicted by Eq.~(\ref{intro:em_asymp}), and for fixed $r_0/M$
  the self-force difference is seen to decrease with increasing $n$;
  this behaviour is discussed and explained in Sec.~\ref{sec:intro}.}  
\label{fig:em_force}
\end{figure}

The self-force difference of Eq.~(\ref{em_force}) can be computed as a  
function of $r_0$ for selected stellar models characterized by a 
polytropic index $n$ and a relativistic parameter $b$. Sample results
are presented in Fig.~\ref{fig:em_force}, and our findings were
discussed more fully in Sec.~\ref{sec:intro}. As in the scalar case the
self-force for a massive thin shell can be computed by inserting 
$\eta_\ell^1 = \ell/\sqrt{1-2M/R}$ within
Eq.~(\ref{scalar_structure}); this reproduces the results of Burko,
Liu, and Soen \cite{burko-etal:01}.  

\ack

This work was supported by the Natural Sciences and Engineering
Research Council of Canada, and by the Grant-in-Aid for the Global COE
Program ``The Next Generation of Physics, Spun from Universality and
Emergence'' from the Ministry of Education, Culture, Sports, Science
and Technology of Japan. S.\ Isoyama is grateful to the Yukawa
Institute for Theoretical Physics at Kyoto University for supporting
his stay at University of Guelph. We thank Takahiro Tanaka for useful
conversations that helped improve the paper.  

\section*{References}
\bibliography{../bib/master} 
\end{document}